\documentclass[10pt,amsmath,amssymb,twocolumn,superscriptaddress,superscriptaddress,floatfix,nofootinbib,aps]{revtex4-1}

\usepackage{graphicx}
\usepackage{dcolumn}
\usepackage{bm}
\usepackage[normalem]{ulem}
\usepackage{setspace}
\usepackage{amsmath}
\usepackage{enumerate}
\usepackage{tabularx}
\usepackage{wrapfig}
\usepackage{ifthen}
\usepackage[utf8]{inputenc}

\usepackage[colorlinks=true]{hyperref}

\def\simge{\mathrel{
     \rlap{\raise 0.511ex \hbox{$>$}}{\lower 0.511ex \hbox{$\sim$}}}}
\def\simle{\mathrel{
     \rlap{\raise 0.511ex \hbox{$<$}}{\lower 0.511ex \hbox{$\sim$}}}}
\def\be{\begin{equation}}
\def\ee{\end{equation}}

\newcommand{\ev}{\, {\rm eV}}

\newcommand{\lr}[1]{ \left( #1 \right) }

\newcommand{\Tr}{ {\rm Tr} \, }

\newcommand{\calz}{{\cal Z}}

\newcommand{\beq}{\begin{eqnarray}}
\newcommand{\eeq}{\end{eqnarray}}

\graphicspath{{./figures/}}

\begin{document}
\sloppy

\title{Numerical evidence  of conformal phase transition in graphene with long-range interactions}

\author{Pavel~Buividovich}
\email{Pavel.Buividovich@physik.uni-regensburg.de}
\affiliation{Institut f\"ur Theoretische Physik, Justus-Liebig-Universit\"at, 35392 Giessen, Germany}
\affiliation{Institut f\"ur Theoretische Physik, Universit\"at Regensburg, 93053 Regensburg, Germany}

\author{Dominik~Smith}
\email{Dominik.D.Smith@theo.physik.uni-giessen.de}
\affiliation{Institut f\"ur Theoretische Physik, Justus-Liebig-Universit\"at, 35392 Giessen, Germany}

\author{Maksim~Ulybyshev}
\email{Maksim.Ulybyshev@physik.uni-wuerzburg.de}
\affiliation{Institut f\"ur Theoretische Physik, Julius-Maximilians-Universit\"at,
   97074 W\"urzburg, Germany}

\author{Lorenz von Smekal}
\email{Lorenz.Smekal@physik.uni-giessen.de}
\affiliation{Institut f\"ur Theoretische Physik, Justus-Liebig-Universit\"at, 35392 Giessen, Germany}

\date{\today}
\begin{abstract}
\noindent 
Using state of the art Hybrid-Monte-Carlo (HMC) simulations we carry out an unbiased study of the competition between spin-density wave (SDW) and charge-density wave (CDW) order in suspended graphene. We determine that the realistic inter-electron potential of graphene must be scaled up by a factor of roughly $1.6$ to induce a semimetal-SDW phase transition and find no evidence for CDW order. A study of critical properties suggests that the universality class of the three-dimensional chiral Heisenberg Gross-Neveu model with two fermion flavors, predicted by renormalization group studies and strong-coupling expansion, is unlikely to apply to this  transition. We propose that our results instead favor an interpretation in terms of a conformal phase transition. In addition, we describe a variant of the HMC algorithm which
uses exact fermionic forces during molecular dynamics trajectories and avoids the use of pseudofermions. 
Compared to standard HMC this allows for a substantial increase of the integrator stepsize while achieving comparable Metropolis acceptance rates and leads to a sizable performance improvement.

\end{abstract}
\maketitle

\section{Introduction}
\label{sec:Intro}

Graphene, with its orders of magnitude higher charge carrier mobility, is considered silicon’s ideal replacement for semiconductor-based devices. However, clean suspended graphene, which features the maximal carrier mobility at the same time lacks an energy gap in its band structure, the existence of which is the prerequisite for building graphene-based transistors. 

Hypothetically such a gap could exist, since the small Fermi velocity of $v_F \approx c/300$ (where $c$ is the speed of light) leads to strong interactions between electronic quasi-particles with an effective fine-structure constant of $\alpha_{\textrm{eff}}=e^2/(\hbar v_F) \approx 2.2$. Thus one expects, based on the analogy to chiral symmetry breaking in quantum field theories, numerous theoretical arguments \cite{Gamayun:2010,KHVESHCHENKO2004323,Araki2010,ARAKI20111408,Araki2012} and numerical simulations \cite{Lahde:09:1,Lahde:09:2,Hands:08:1,Hands:10:1,Buividovich:12:1}, that for $\alpha_{\textrm{eff}}$ larger than some critical value $\alpha_c \approx 1$, interactions destabilize the system towards spontaneous formation of gapped ordered phases. Besides the well-known anti-ferromagnetic spin-density wave (SDW) order favored by sufficiently strong on-site repulsion \cite{Herbut2006,Herbut2009,Herbut2009:2,Semenoff2012}, various combinations of nearest neighbor and other short-range couplings might also induce such phases as a charge-density wave (CDW) \cite{Sorella:1992,Semenoff:84:1,Herbut:cond-mat/0606195,Semenoff:1204.4531,Gracey:1801.01320}, topological insulators \cite{Raghu:07:1}, spontaneous Kekul\'e distortions \cite{Mudry:cond-mat/0609740,PhysRevB.96.115132} as well as coupled spin-charge-density-wave phases \cite{Makogon:1007.0782} and spin spirals \cite{PhysRevB.70.195122}.
Also the existence of triple or multicritical points at which semimetal, CDW and SDW phases meet has been discussed \cite{Herbut:cond-mat/0606195,PhysRevB.92.035429,PhysRevB.93.125119}.

Experimentally on the other hand, it has been firmly established that suspended graphene is a semimetal \cite{Elias:2011,Elias:12:1}. This implies that electronic two-body interactions are still too weak to induce a semimetal-insulator quantum phase transition. The absence of an energy gap has been reproduced in first-principle numerical simulations \cite{Buividovich:13:5,Smekal:1403.3620,PhysRevB.94.085421} which properly take into account the screening of the bare Coulomb potential by electrons in lower $\sigma$-orbitals \cite{Wehling:1101.4007}. This screening increases the critical coupling for a semimetal-insulator transition up to roughly $\alpha_c\approx 3.1$, which is noticeably higher than the effective coupling strength $\alpha_\textrm{eff} \approx 2.2$ in suspended graphene. Similarly, numerical calculations of the conductivity yielded a finite result almost equal to that of non-interacting graphene, implying the absence of a band gap \cite{PhysRevB.94.085421}.

Despite being in the weak-coupling gapless regime, suspended graphene may still be quite close to a semimetal-insulator transition. The knowledge of how close real graphene is to a phase transition might still help to describe the strong-coupling aspects of the many-body physics of this material. An obvious example of the relevance of the position and order of the closest phase transition in the weak-coupling regime are the convergence radius and rate of the perturbative expansion. On the experimental side, applying mechanical strain can move suspended graphene closer to the phase transition \cite{Assaad:1505.04188,PhysRevB.96.155114}.

Guided by the results obtained within renormalization group techniques \cite{Herbut2006,Herbut2009,Herbut2009:2} and strong-coupling expansion \cite{Semenoff2012}, most numerical studies \cite{Buividovich:13:5,Smekal:1403.3620,Assaad:1505.04188} have focused on detecting the onset of spin-density wave (SDW) order, which is expected to be a second-order phase transition in the universality class of the $N_f=2$ chiral Heisenberg Gross-Neveu model in three space-time dimensions.\footnote{The chiral Gross-Neveu universality class has been verified for the hexagonal Hubbard model with purely on-site interactions through numerous numerical studies. See e.g. Refs. \cite{Assaad:1304.6340,PhysRevX.6.011029,PhysRevB.91.165108,PhysRevB.90.085146}.} 
Within the perturbative renormalization-group analysis the robustness of this scenario is supported by the observation that the long-range Coulomb potential is a marginally irrelevant interaction \cite{Herbut2006,Herbut2009,Herbut2009:2}, and thus the semimetal-insulator phase transition should be driven by on-site interactions.

Since the screening of the bare Coulomb potential by electrons in $\sigma$-orbitals mostly suppresses short-distance interactions and the long-range potential is only weakly affected \cite{Wehling:1101.4007,Assaad:1505.04188}, the long-range Coulomb interaction might still dominate the near-critical behavior \cite{Herbut2009:2,Gamayun:2010}. This might favor ordered phases other than the anti-ferromagnetic SDW phase, with the charge-density wave being the most likely candidate \cite{Sorella:1992,Semenoff:84:1,Herbut:cond-mat/0606195,Semenoff:1204.4531,Gracey:1801.01320}.  An unbiased study from first principles of the competition between different ordered phases in the vicinity of suspended graphene, considered as a point in the space of all possible inter-electron interactions, is thus desirable. 

By extension, the universality class (and even the order) of the possible semimetal-insulator transition also remains unclear. At present, the only first-principles calculations of critical exponents were carried out in the Dirac cone approximation \cite{Lahde:09:1,Lahde:09:2,Hands:08:1,Hands:10:1} and have not unambiguously settled the issue. The prediction of Gross-Neveu scaling relies upon an identification with the Hubbard model with on-site interactions only, which is a rather drastic modification. 

A transition to a phase other than SDW could certainly imply different critical properties: It has been argued, for instance, that a semimetal-CDW transition should fall into the chiral Ising universality class \cite{Janssen2014}. 
Another interesting possibility is that the scale-invariant $\sim 1/r$ Coulomb interaction induces the so-called conformal phase transition (CPT) of infinite order \cite{PhysRevD.80.125005}, at which physical observables exhibit exponential (``Miransky'') rather than powerlaw scaling \cite{Miransky1997}. CPT generalizes the concept of the Berezinskii-Kosterlitz-Thouless transition \cite{Berezinskii1970,Kosterlitz1973} to higher than two dimensions where long-range order is possible. It is a continuous transition characterized by an exponentially increasing 
correlation length and occurs when changes of some control parameter cause the merging of infrared and ultraviolet renormalization group fixed points which marks the transition from a conformal to a non-conformal phase \cite{PhysRevD.80.125005}. For graphene modelled as $2+1$-dimensional Dirac fermions with bare Coulomb interaction, a CPT is predicted by the analysis based on Schwinger-Dyson equations \cite{PhysRevB.66.045108, KHVESHCHENKO2004323}.

In this work, we use our state of the art Hybrid Monte-Carlo simulation code with numerous improvements discussed recently in Ref. \cite{Buividovich:2018hubb} to check how close suspended graphene might be to a semimetal-insulator transition, and to address the properties of this transition. We use a realistic partially screened Coulomb potential \cite{Smekal:1403.3620} which accounts for screening by electrons in lower $\sigma$-orbitals \cite{Wehling:1101.4007} and drive the system towards the transition by uniformly rescaling this potential, as in Ref. \cite{Buividovich:13:5}. We thus improve and re-check the results of previous studies \cite{Buividovich:13:5,Smekal:1403.3620,Assaad:1505.04188} which might have significant systematic errors due to small lattice sizes, large discretization artifacts, and artificial mass terms. The most important improvements include:
\begin{itemize}
 \item Unlike in Refs. \cite{Buividovich:13:5,Smekal:1403.3620} we study the competition of SDW and CDW phases in a completely unbiased way, without symmetry-breaking mass terms that favor a specific phase. To this end we use quadratic observables as order parameters, such as the squared spin or charge per sublattice \cite{Buividovich:2018hubb}, which unlike the corresponding condensates are non-zero in finite volume even without external sources and allow for an unambiguous determination of the ground state. 
 \item To avoid the loss of ergodicity of HMC simulations caused by zero modes of the tight-binding Hamiltonian without mass terms, we represent inter-electron interactions in terms of complex-valued Hubbard-Stratonovich fields \cite{Assaad:1708.03661,Ulybyshev:1712.02188,Buividovich:2018hubb}.
 \item An improved fermionic lattice action with exact sublattice (chiral) symmetry and strongly suppressed discretization errors \cite{Buividovich:2018hubb,Buividovich:16:4}.
 \item An efficient non-iterative Schur complement solver which significantly speeds up the simulations \cite{Ulybyshev:2018dal}.
 \item {Molecular dynamics trajectories which use exact fermionic forces and avoid the use of pseudofermions, which leads to another sizable performance improvement. This is a very recent development and is described in detail in Section \ref{subsec:hmc_exact_forces}.}
 \item Several improvements of the simulation parameters such as lower electronic temperatures ($T=0.125 \ev$ instead of $0.5 \ev$), larger spatial lattice sizes (up to $L=24$ instead of $18$) and finer discretization of the Euclidean time axis ($\delta_{\tau}=0.0625 \ev^{-1}$ instead of $0.1 \ev^{-1}$).
\end{itemize}

Using infinite-volume extrapolations of order parameters, we are able to determine that  SDW order spontaneously forms, without being favored by a source term, at a critical coupling of $\alpha_c\approx 3.5$. This is larger than the previous estimate $\alpha_c\approx 3.1$ and implies that the scenario of suspended graphene being in the semimetal phase remains stable under our improvements of the HMC method. With high confidence we also rule out the presence of CDW order for couplings of $\alpha_\textrm{eff} \lesssim 5$. 

Furthermore, we address the question of the universal properties of the phase transition induced by rescaling the screened inter-electron interaction potential in suspended graphene \cite{Wehling:1101.4007}. Quite intriguingly, we find indications that the ratio between on-site and non-local interactions in the screened potential might favour the infinite-order phase transition scenario predicted in Refs.
\cite{PhysRevB.66.045108, KHVESHCHENKO2004323}. 

By studying the finite-size scaling of the squared spin per sublattice, we find that the ratio of critical exponents $\beta/\nu$ is close to exactly one in good approximation for the semimetal-SDW transition. Furthermore, the collapse of data points from different lattice sizes onto a universal scaling function is rather insensitive toward the choice of correlation length exponent $\nu$, with the optimal choice drifting slightly towards larger values when smaller lattices are excluded from the analysis. We obtain evidence that a collapse may occur naturally in infinite volume, without the need for a rescaling factor $L^{1/\nu}$, which formally corresponds to the limit $\nu \to \infty$. We argue that such behavior is consistent with a phase transition governed by Miransky scaling, with finite-volume corrections which mimic a second order phase transition on small systems. We also compare our data with reference data obtained for the Hubbard model with purely on-site interactions, and conclude that the interpretation of the numerical results in terms of Gross-Neveu scaling is much less convincing for our non-local interaction potential than for purely on-site interactions.

\section{Simulation setup}
\label{sec:Setup}

\subsection{The path-integral formulation of the partition function}
\label{subsec:hmc}

The basic idea of first-principle Monte-Carlo simulations is to carry out a stochastic integration of the functional integral representation of the grand-canonical partition function $\calz=\Tr e^{-\beta \hat{\mathcal{H}}}$, in which operators are replaced by fields, by using a Markov process which evolves the fields in computer time such that their time histories are consistent with the equilibrium distribution. Thermodynamic expectation values of observables $\langle \hat{O}\rangle= \frac{1}{\calz}\Tr(\hat{O}e^{-\beta \hat{\mathcal{H}}})$ are then obtained from  measurements on a representative set of field configurations. 

Our starting point is the interacting tight-binding Hamiltonian on the hexagonal lattice, in second quantized form:
\beq 
\label{eq:tightbinding}
\hat{\mathcal{H}} = - \kappa \sum_{\left \langle x, y \right \rangle, \sigma} ( \hat{c}^{\dagger}_{x,\sigma} \hat{c}_{y,\sigma} + \text{h.c.}) + \frac{\lambda}{2}\sum_{x,y}  \hat{\rho}_x V_{xy}\hat{\rho}_y .
\eeq
Here $\kappa$ is the hopping parameter, $\langle x, y\rangle$ denotes nearest-neighbor sites, $\sigma = \uparrow, \downarrow$ labels spin components and $\hat{\rho}_x =\hat{c}^{\dagger}_{x,\uparrow} \hat{c}_{x,\uparrow}+\hat{c}^{\dagger}_{x,\downarrow} \hat{c}_{x,\downarrow}-1$ is the electric charge operator.
The creation- and annihilation operators satisfy the anticommutation relations $\{ \hat{c}_{x,\sigma}, \hat{c}^{\dagger}_{y,\sigma'} \}= \delta_{x,y} \delta_{\sigma, \sigma'}$. $V_{xy}$ is the partially screened Coulomb potential used in Refs. \cite{Smekal:1403.3620, Korner:2017qhf}. To drive the system towards the semimetal-insulator phase transition, we rescale this potential by a factor $\lambda > 1$, so that suspended graphene corresponds to $\lambda=1$. As the cRPA data of \cite{Assaad:1505.04188} suggests, for not very large distances of order of few lattice spacings, the effect of strain can be roughly described in terms of such a rescaling.

\begin{figure}[h!tpb]
  \includegraphics[width=0.48\textwidth]{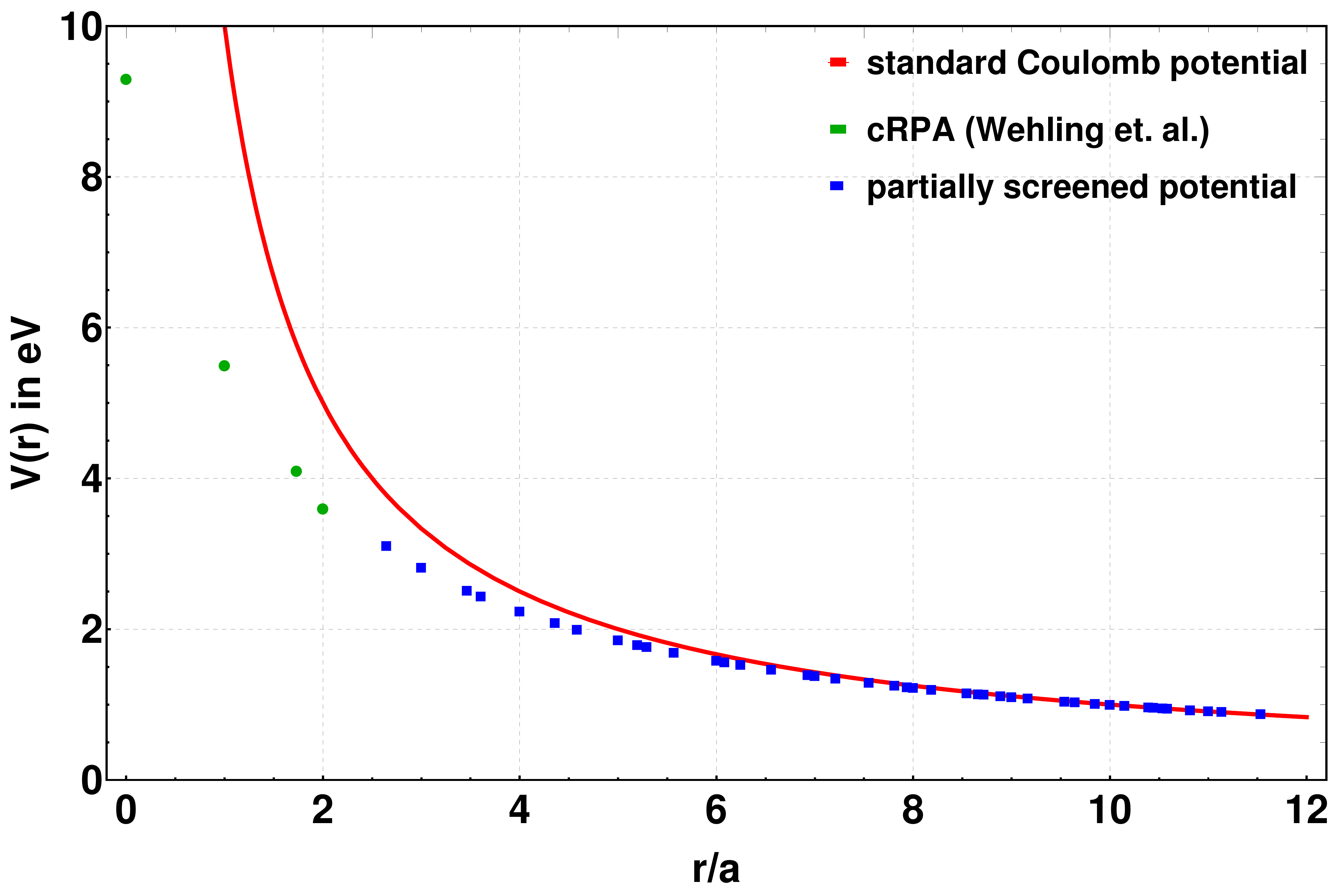}\\
\caption{Partially screened inter-electron interaction potential given by (\ref{eq:potfit}) (green and blue) compared with the bare Coulomb potential (red). Green and blue points correspond to the cRPA values (green) and to the thin-film model potential (blue), respectively, all taken from \cite{Wehling:1101.4007}.}  
\label{fig:potential}    
\end{figure}

The potential $V_{xy}$ contains the exact values obtained from calculations within a constrained random-phase approximation (cRPA) by Wehling \textit{et al.}~in Ref.~\cite{Wehling:1101.4007} for the on-site $U_{00}$, nearest-neighbor $U_{01}$, next-nearest-neighbor $U_{02}$, and third-nearest-neighbor $U_{03}$ interaction parameters.\footnote{There is still some minor disagreement over the exact values of these parameters in graphene \cite{Assaad:1505.04188}. The uncertainties are most likely too small to have any significant effect on the results of this work however.} At longer distances a momentum dependent phenomenological dielectric screening formula, derived also in Ref.~\cite{Wehling:1101.4007} based on a thin-film model, is used  for a smooth interpolation to an unscreened Coulomb tail. Both results are combined via a parametrization based on a distance dependent Debye mass $m_D$. The matrix elements $V_{xy}$ are then filled using   
\begin{equation}
   V(r)  = \left\{ \begin{array}{lr}      
     U_{00},U_{01},U_{02},U_{03} & , \; r\le 2a \\[2pt]
     e^2  \left( c \, \displaystyle
     \frac{\exp(-m_D r)}{a (r/a)^{\gamma}}  + m_0
   \right)  &, \; r > 2a  
      \end{array}
          \right.\label{eq:potfit}
\end{equation}
where $a$ is the nearest-neighbor distance and  $m_D$,
$m_0$, $c$ and $\gamma$ are piecewise constant chosen such that $m_D,
m_0 \to 0$ and $c, \gamma \to 1$ for $r \gg a $. Tables with precise
values of these parameters can be found in Ref.~\cite{Smekal:1403.3620}.
Fig.~\ref{fig:potential} shows this interaction potential in
comparison to the unscreened Coulomb potential.   

To avoid a fermion sign problem (where the measure of the functional integral becomes complex or of indefinite sign, which prevents importance sampling), we apply the following canonical transformation to the Hamiltonian (\ref{eq:tightbinding}): Hole creation and annihilation operators $\hat{b}^\dagger_x,\hat{b}_x$ are introduced for the spin-down electrons and the sign of these is then flipped on one of the triagonal sublattices of the hexagonal lattice. The transformation law can be summarized as
\begin{eqnarray}
\hat{c}_{x, \uparrow}, \hat{c}^{\dagger}_{x, \uparrow} &\to & \hat{a}_x, \hat{a}^{\dagger}_x,
\nonumber \\ 
\label{eq:trafo1}
\hat{c}_{x, \downarrow}, \hat{c}^{\dagger}_{x, \downarrow} &\to & \pm \hat{b}^{\dagger}_x, \pm \hat{b}_x,
\end{eqnarray}
where the signs in the second line alternate between the two sublattices. This leads to $\hat{\rho}_x = \hat{a}^{\dagger}_x \hat{a}_x - \hat{b}^{\dagger}_x \hat{b}_x$.
We also apply the following Fierz transformation to the on-site interaction term:
\begin{align}
\label{eq:complexint}
\frac{V_{xx}}{2} \hat{\rho}_x^2 =&
\eta \frac{V_{xx}}{2} \hat{\rho}_x^2 - (1-\eta) \frac{V_{xx}}{2}(\hat{\rho}_x')^2\notag \\ & + V_{xx}(1-\eta)\,\hat{\rho}_x',
\end{align} Here $\hat{\rho}_x' = \hat{a}^{\dagger}_x \hat{a}_x + \hat{b}^{\dagger}_x \hat{b}_x$ is the spin-density operator and the constant $\eta$ can be chosen in the range $(0,1)$. The purpose of this transformation is to extend the Hubbard fields (introduced below) to complex numbers. This is necessary
when the Hamiltonian (\ref{eq:tightbinding}) contains no mass terms,
as the configuration spaces of both purely real and purely imaginary auxiliary fields then form disconnected regions, separated by infinitely high potential barriers (extended manifolds where the fermion determinant vanishes). The additional degrees of freedom of complex fields allow our Monte-Carlo algorithm to circumvent the barriers and ensure ergodicity \cite{Assaad:1708.03661,Ulybyshev:1712.02188,Buividovich:2018hubb}. The constant $\eta$ interpolates between real and imaginary fields.

To derive the functional integral, we start with a symmetric Suzuki-Trotter decomposition  which yields 
\begin{eqnarray}
\calz &\approx &\Tr \left(
\prod_{i=1}^{N_\tau}
e^{-\delta_{\tau}(\hat{\mathcal{H}}_{0} +\mathcal{H}_{\text{int}})} \right) \notag\\ &=&
\Tr \left( e^{-\delta_{\tau} \hat{\mathcal{H}}_{0}} e^{-\delta_{\tau} \hat{\mathcal{H}}_{\text{int}}} e^{-\delta_{\tau} \hat{\mathcal{H}}_{0}} \dots \right) + O(\delta^2_{\tau}),
\label{eq:Trotter}
\end{eqnarray}
where the exponential is factorized into $N_\tau$ terms and the kinetic $\hat{\mathcal{H}}_{0}$ and interaction $\hat{\mathcal{H}}_{\text{int}}$ contributions are separated. This introduces a finite step size $\delta_{\tau} = \beta/N_{\tau}$ in Euclidean time and a discretization error $O(\delta^2_{\tau})$. The four-fermion terms appearing in $\hat{\mathcal{H}}_{\text{int}}$ are now converted into bilinears by Hubbard-Stratonovich (HS) transformation. We use two distinct variants: The first term $\sim \eta \hat{\rho}_x^2 $ appearing on the right hand side of Eq. (\ref{eq:complexint}) is re-absored into the interaction matrix $V_{xy}$ appearing in  (\ref{eq:tightbinding}) and the combined expression is then transformed using
\begin{eqnarray}
\label{continuous_HS_imag}
  e^{-\frac{\delta_\tau}{2}
  \underset{x,y}{\sum}
   V_{xy} \hat \rho_x \hat \rho_y} \cong \int D \phi \,
  e^{- \frac{1}{2\delta_\tau} \underset{x,y}{\sum} \phi_x V^{-1}_{xy} \phi_y} e^{i \underset{x}{\sum} \phi_x \hat \rho_x}.
\end{eqnarray}
The  $\sim (1-\eta)(\hat{\rho}_x')^2$ term in Eq. (\ref{eq:complexint}) is transformed by its own, using
\begin{eqnarray}
 \label{continuous_HS_real}
   e^{\frac{\delta_\tau}{2}(1-\eta) \underset{x}{\sum} V_{xx} (\hat \rho_x')^2 } \cong \int D \chi\, e^{- \frac{1}{2\delta_\tau} \underset{x}{\sum} \frac{\chi^2_x}{(1-\eta)V_{xx}} } e^{ \underset{x}{\sum} \chi_x \hat \rho_x'}.
\end{eqnarray}
In effect, we have introduced a complex bosonic auxiliary field $\Phi$ (``Hubbard field'') with real part $\chi$ and imaginary part $i\phi$. Note that the transformations are applied once to each timeslice, leading to $\phi\equiv\phi_{x,t}$ and $\chi\equiv\chi_{x,t}$. The third term  in Eq. (\ref{eq:complexint}) is already a bilinear and doesn't need to be transformed. Due to the translational invariance of the integration measure in Eq. (\ref{continuous_HS_real}) it can be absorbed into the real part of the Hubbard field through the transformation 
\begin{eqnarray}
\label{eq:shiftdiag}
\chi \to\chi-\delta_\tau V_{xx}(1-\eta).
\end{eqnarray}

To compute the trace in the fermionic Fock space (with anti-periodic boundary conditions) appearing in Eq. (\ref{eq:Trotter}) we use
\begin{align}
\label{fermionic_identity}
 &\Tr\left( e^{-\hat{A}_1 } e^{-\hat{A}_2 } \ldots e^{-\hat{A}_n } \right)
 = \nonumber \\ &=
 \det\left(
  \begin{array}{cccc}
     1          & -e^{-A_1} & 0        & \ldots  \\
     0          & 1        & -e^{-A_2} & \ldots  \\
        \vdots  &          & \ddots          &         \\
      e^{-A_n} & 0        & \ldots   & 1       \\
  \end{array}
 \right),
\end{align}
where $\hat{A}_k = \lr{A_k}_{ij} \hat{c}^{\dag}_i \hat{c}_j$ are the fermionic bilinear operators and $A_k$ (without hat) contain matrix elements in the single-particle Hilbert space. This identity is derived in Refs.~\cite{Hirsch:85:1,Blankenbecler:81:1,MontvayMuenster}. Applying (\ref{fermionic_identity}) to Eq. (\ref{eq:Trotter}), we obtain
\begin{equation}
\label{eq:func_int}
  \calz = \int D \Phi \,|\det M( \Phi)|^2
  e^{-S_\eta(\Phi)},\,
  \end{equation}
  with
  \begin{align}
  S_\eta(\Phi) =& \frac{1}{ 2  \delta_\tau}\sum_{x,y,t} \phi_{x,t}\widetilde{V}_{xy}^{-1}\phi_{y,t}
  \notag\\&+ \sum_{x,t} \frac {(\chi_{x,t}- (1-\eta) \delta_\tau V_{xx})^2} {2 (1-\eta) \delta_\tau V_{xx}}.
  \label{eq:action_alpha}
\end{align}
Here $\widetilde{V}$ denotes a modified interaction matrix wherein the diagonal elements have been rescaled by a factor of $\eta$ by the Fierz transformation (\ref{eq:complexint}). The constant shift of $\chi$ in the second sum results from Eq. (\ref{eq:shiftdiag}). The fermion matrix is given by 
\begin{align}
\label{eq:fermionmatrix}
 &M(\Phi) 
 = \nonumber \\ &=
 \left(
  \begin{array}{cccccc}
     1          & -e^{-\delta_\tau h} & 0        & 0 & 0 &\ldots  \\
     0          & 1        & -e^{i \Phi_{1}} & 0 & 0 & \ldots  \\
     0          & 0        & 1                & -e^{-\delta_\tau h} & 0 &    \ldots\\
     0          & 0 & 0     & 1                & -e^{ i \Phi_{2}} &    \ldots\\
         \vdots & & & & \ddots &                               \\
     e^{i \Phi_{N_\tau}} & 0  & 0 &       & \ldots   & 1       \\
  \end{array}
 \right) ,
\end{align}
where we use the short-hand notation 
$e^{i \Phi_t} \equiv \textrm{diag}\left( e^{\chi_{x,t}+i \phi_{x,t}} \right)$ and $h$ denotes the single-particle tight-binding hopping matrix. $|\det M( \phi)|^2$ appears in (\ref{eq:func_int}) since  the fermionic matrices for spin-up and spin-down electrons are $M$ and $M^\dagger$, respectively. 

Eq.  (\ref{eq:func_int}) is exactly of the form required for Monte-Carlo simulations: $\calz$ is expressed as a functional integral over classical field variables with a positive-definite measure. We point our here that the appearence of matrix exponentials $e^{-\delta_\tau h}$ in the fermion matrix $M(\Phi)$ leads to a non-local fermion action. This action has an exact sublattice-particle-hole symmetry, even at finite $\delta_\tau$ and in the presence of the fluctuating Hubbard fields \cite{Buividovich:2018hubb,Buividovich:16:4}. In contrast, the linearized action used in the previous studies \cite{Smekal:1403.3620,Korner:2017qhf,Buividovich:13:5} corresponds to expanding the blocks $e^{-\delta_\tau h}$ in $M(\Phi)$ to linear order in $\delta_\tau$. The main disadvantage of this linearized formulation is that the leading discretization errors generate a strong explicit breaking of the spin rotational symmetry in this case, which is only suppressed at very large $N_\tau$  \cite{Buividovich:2018hubb,Buividovich:16:4}. $M(\Phi)$ is a dense matrix here, which makes iterative inversion methods such as the standard conjugate-gradient solver rather inefficient. $M(\Phi)$ can be efficiently inverted however, using a recently developed solver based on Schur decomposition \cite{Ulybyshev:2018dal}.

\subsection{Hybrid-Monte-Carlo with exact fermionic forces}
\label{subsec:hmc_exact_forces}

This work employs the Hybrid-Monte-Carlo (HMC) algorithm based on the formalism originally developed in Refs.~\cite{Rebbi:11:1,Rebbi:12:1} to study the graphene tight-binding model with interactions. HMC has its origins in lattice QCD simulations \cite{DeGrandDeTarLQCD,MontvayMuenster,Buividovich:16:1} but is increasingly being applied also in condensed matter physics \cite{Hands:08:1,Hands:10:1,Hands:11:1,DelDebbio:96:1,Lahde:09:1,Lahde:09:2,Lahde:09:3,Lahde:11:1,Rebbi:12:1,Buividovich:13:5,Smekal:1403.3620,Assaad:1708.03661,Ulybyshev:2015opa,Ulybyshev:2017szp,PhysRevB.94.085421,PhysRevB.95.165442,PhysRevB.94.245112,Luu:2015gpl,Berkowitz:2017bsn,PhysRevB.96.165411,PhysRevB.96.205115} alongside determinantal Quantum-Monte-Carlo simulations following Blankenbecler, Scalapino and Sugar (BSS-QMC) \cite{Blankenbecler:81:1,Scalettar:86:1}. As we have described the individual steps of HMC in detail in several publications \cite{Smekal:1403.3620,Korner:2017qhf,Buividovich:2018hubb} we will focus entirely on a recent development here, whereby the algorithm is implemented with exact fermionic forces rather than using pseudofermions.

The HMC algorithm includes molecular dynamics (MD) trajectories, during which the Hubbard field is evolved in computer time by an artificial Hamiltonian process. During these trajectories the effective action
\begin{eqnarray}
\label{eq:eff_action}
 S_{\textrm{eff.}}(\Phi) =  S_\eta(\Phi) - \ln ( \det M( \Phi) \det M( \Phi)^\dag) 
 \end{eqnarray}
plays the role of potential energy for the Hubbard field $\Phi$. Obviously, one needs to compute the derivative of the effective action with respect to Hubbard field in order to solve Hamilton's equations. The standard approach is to use a stochastic representation of the determinants in Eq. (\ref{eq:eff_action}):
\begin{eqnarray}
\label{eq:stoch_det}
|\det M( \Phi)|^2 = \int D \Psi \Psi^\dag e^{-\Psi^\dag (M M^\dag)^{-1}\Psi },
 \end{eqnarray}
which introduces an additional pseudofermionic field $\Psi$. Calculations of derivatives of $S_{\textrm{eff.}}(\Phi)$ with respect to $\Phi$ then require just one solution of the linear equation $M M^\dag \Psi = X$, where $X$ is a Gaussian distributed field. This solution can be obtained using an iterative solver or a non-iterative solver \cite{Ulybyshev:2018dal}. The latter strategy was used in Ref. \cite{Buividovich:2018hubb}.

However, one can go even further and avoid pseudofermions entirely, by computing the derivatives of $S_{\textrm{eff.}}(\Phi)$ directly, starting from Eq. (\ref{eq:eff_action}). Calculations of derivatives of $S_\eta(\Phi)$ are trivial, while derivatives of
$\ln ( \det M( \Phi) \det M( \Phi)^\dag)$
can be computed using:
\begin{eqnarray}
\label{eq:det_der}
 \frac{\partial \ln \det M} {\partial \phi_{x,t}} =  \Tr\left( { M^{-1} \frac {\partial M}{\partial \phi_{x,t}}  }\right).
 \end{eqnarray}
 It turns out that this requires the knowledge of only a few elements of the fermion propagator $M^{-1}$. Due to the special band structure of the matrix $M$ given by Eq. (\ref{eq:fermionmatrix}), we need only elements of $M^{-1}$ which are located in blocks immediately off the main diagonal. 
 
To proceed, let us write the fermionic operator (\ref{eq:fermionmatrix}) in the general form:
\begin{align}
\label{eq:fermionmatrix_general}
 &M(\Phi) 
 = \nonumber \\ &=
 \left(
  \begin{array}{cccccc}
     1          & D_1 & 0        & 0 & 0 &\ldots  \\
     0          & 1        & D_2 & 0 & 0 & \ldots  \\
     0          & 0        & 1                & D_3 & 0 &    \ldots\\
     0          & 0 & 0     & 1                & D_4 &    \ldots\\
         \vdots & & & & \ddots &                               \\
     D_{2 N_\tau} & 0  & 0 &       & \ldots   & 1       \\
  \end{array}
 \right) ,
\end{align}
where even blocks $D_{2k}$ with $k=1...N_\tau$ correspond to diagonal matrices containing the exponentials $\pm e^{i\phi_{x,t}+\chi_{x,t}}$, and odd blocks are equal to exponentials of the tight-binding Hamiltonian $-e^{-\delta_\tau h}$. The inverse fermionic matrix can also be written in terms of spatial blocks:
\begin{align}
\label{eq:propagator_general}
 &M^{-1}(\Phi) 
 = \nonumber \\ &=
 \left(
  \begin{array}{cccccc}
     g_1          & \ldots & \ldots        & \ldots & \ldots & \bar g_{2 N_\tau}  \\
     \bar g_1          & g_2        & \ldots & \ldots & \ldots & \ldots  \\
     \ldots         & \bar g_2        & g_3                & \ldots & \ldots &    \ldots\\
     \ldots          & \ldots & \bar g_3     & g_4                & \ldots &    \ldots\\
         \vdots & & & & \ddots &                               \\
     \ldots & \ldots  & \ldots &       & \ldots   & g_{2 N_\tau}       \\
  \end{array}
 \right).
\end{align}
The matrix $M^{-1}$ is dense, but here we explicitly show only those blocks which are needed for our calculations. In fact, in the trace in Eq.~(\ref{eq:det_der}) only the even blocks $\bar g_{2k}$ for all $k=1...N_\tau$ will contribute to the exact derivatives for computing the fermionic force. 

We can now use part of the BSS-QMC algorithm \cite{Scalettar:86:1} to compute the desired blocks of the propagator. Due to the structure of $M$, the diagonal blocks of $M^{-1}$ can be formally written as 
\begin{eqnarray}
\label{eq:g_diag}
 g_i=\left( { I+D_i D_{i+1} ... D_{2 N_\tau} D_1 ... D_{i-1}  } \right)^{-1},
 \end{eqnarray}
and the following iteration formula can be proven:
\begin{eqnarray}
\label{eq:g_diag_iter}
 g_{i+1}= D_i^{-1} g_i D_i.
\end{eqnarray}
Analogously, off-diagonal blocks of $M^{-1}$ can be written as 
\begin{eqnarray}
\label{eq:g_off_diag}
 \bar g_i= D_{i+1} ... D_{2 N_\tau} g_i~, \end{eqnarray}
 which leads to the relation:
 \begin{eqnarray}
\label{eq:g_off_diag_iter}
 \bar g_{i+1}= D_{i+1}^{-1} \bar g_i D_i.
 \end{eqnarray}
 We can now either directly use Eq. (\ref{eq:g_off_diag_iter}) to obtain the $\bar g_i$ or first obtain the $g_i$ and use the relation 
  \begin{eqnarray}
\label{eq:g_off_diag_g_diag}
 \bar g_{i}= D_{i}^{-1} (I-g_i)~,
 \end{eqnarray}
 between diagonal and off-diagonal blocks.
By iterating either (\ref{eq:g_off_diag_iter}) or (\ref{eq:g_diag_iter}) we can easily find all elements of $M^{-1}$ needed for the computation of the derivative, starting from just one block, which is computed from scratch using the Schur complement solver \cite{Ulybyshev:2018dal}. This is done by applying the solver to point sources in the corresponding time slice. 

An important point here is that the whole procedure scales as ${N_S}^3 N_\tau$, where $N_S$ is the number of sites in one Euclidean time slice of the lattice, so the scaling is not worse than that of the Schur complement solver itself. In practice however, the iterations (\ref{eq:g_off_diag_iter}) and (\ref{eq:g_diag_iter}) suffer from the accumulation of round-off errors, which limits the number of times they can be applied (this number depends mostly on the condition number of  $e^{-\delta_\tau h}$). Afterwards, the block of
$M^{-1}$
in the subsequent time slice must be computed from scratch. This is the so-called stabilization which is routinely used in BSS-QMC \cite{SciPostPhys.3.2.013}. 

Finally, an additional simplification comes from the fact that we do not even need the full Schur complement solver for the computation of the blocks $g_i$ or $\bar g_i$. In order to demonstrate this, we sketch the essential parts of the solver. A more detailed description can be found in Ref. \cite{Ulybyshev:2018dal}. 

Essentially, the solver consists of tree stages. In the first stage we decrease the size of the linear system in an iterative procedure. At each step, the system has the form
 \begin{eqnarray}
\label{eq:schur_system}
 \bar M^{(l)} X^{(l)} = Y^{(l)},
 \end{eqnarray}
where $l$ denotes the step number. We start from the initial system with the matrix $M^{(0)} = M$, the unknown vector $X^{(0)}$ containing elements of the fermionic propagator, and a point source vector $Y^{(0)}$. In the simplest case, when $N_\tau$ is some power of 2, the size of the system decreases as $\bar N_\tau^{(l)} = N_\tau/ 2^{l-1}$. The general case is only slightly more complicated and described in Ref. \cite{Ulybyshev:2018dal}. 

The matrix $M^{(l)}$ always has the same form, with unit matrices in the diagonal blocks and with off-diagonal blocks $D^{(l)}_k$ for $k=1...\bar N_\tau^{(l)}$. Iterations are described by the relations
 \begin{eqnarray}
  \label{eq:schur_iterations1_m}
D^{(l+1)}_k &=& -D^{(l)}_{2k} D^{(l)}_{2k+1}, \quad\quad k=1...\bar N_\tau^{(l)} - 1,  \\
D^{(l+1)}_k &=& -D^{(l)}_{2k} D^{(l)}_{1},\quad\quad\quad k=\bar N_\tau^{(l)}, \nonumber 
 \end{eqnarray}
 for matrices and 
 \begin{eqnarray}
\label{eq:schur_iterations1_v}
Y^{(l+1)}_k &=& Y^{(l)}_{2k} - D^{(l)}_{2k} Y^{(l)}_{2k+1}, \quad k=1...\bar N_\tau^{(l)} - 1, \\
Y^{(l+1)}_k &=& Y^{(l)}_{2k} - D^{(l)}_{2k} Y^{(l)}_{1}, \quad\quad k=\bar N_\tau^{(l)} . \nonumber
 \end{eqnarray}
 for vectors. $Y^{(l)}_k$ denotes the $k$-th timeslice of the vector $Y^{(l)}$. 

The second stage is LU decomposition and solution of the compactified system at $l=l_{max}$. Thus we know the vector $X^{(l_{max})}$. Finally, the third stage is the reversed iterative process of reconstruction of the initial solution starting from $X^{(l_{max})}$, using matrix blocks $D^{(l)}_k$ and vectors $Y^{(l)}_{k}$ computed during the first stage:
\begin{eqnarray}
\label{eq:schur_iterations2}
X^{(l)}_{2k} &=& Y^{(l)}_{2k-1} - D^{(l)}_{2k-1} X^{(l+1)}_{k}, \quad k=1...\bar N_\tau^{(l+1)}, \\
X^{(l)}_{2k} &=& X^{(l+1)}_{k}, \quad\quad\quad\quad
\quad\quad\quad\,\,
k=1...\bar N_\tau^{(l+1)}. \nonumber
 \end{eqnarray}
In the end, we arrive at the initial vector $X^{(0)}$  which gives us the matrix elements of the fermionic propagator. 

\begin{figure}[h!tpb]
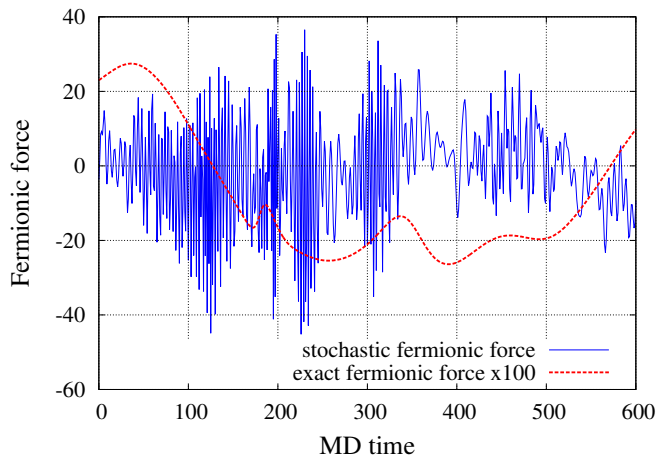

  \includegraphics[width=0.35\textwidth, angle=270]{{{joint_force}}}\\
    \caption{Fermionic forces acting on Hubbard field $\phi_{x,t}$ during a MD trajectory. Example calculation is made for a $6\times6$ hexagonal lattice with electron-electron interaction corresponding to suspended graphene. $N_\tau=128$ and temperature is equal to 0.125 eV. $\partial \ln \det M / \partial \phi_{x,t}$ is shown  for exact fermionic forces and $\partial \Psi^\dag (MM^\dag)^{-1} \Psi / \partial \phi_{x,t}$ is shown for the stochastic representation of the determinant with pseudofermions. Exact forces are rescaled for visibility. }
    \label{fig:joint_force}
\end{figure}

One should note that the initial vector $Y^{(0)}$ contains non-zero elements only in one time slice. Due to the structure of the iterations (\ref{eq:schur_iterations1_v}), this feature is preserved at each step, thus we actually do not need to make the full loop over $k$ in (\ref{eq:schur_iterations1_v}). The same is true for backward iterations  (\ref{eq:schur_iterations2}), for a different reason: we need only one time slice in the final solution $X^{(0)}$, since we are interested either only in diagonal blocks $g_k$ or only in off-diagonal blocks $\bar g_k$. Due to this simplification we need only one matrix-vector operation for each of the few time slices in which we actually recompute the elements of fermionic propagator from scratch. Thus the numerical cost of the method is dominated by matrix-matrix operations (\ref{eq:schur_iterations1_m}) and (\ref{eq:g_off_diag_iter}). This means that the number of floating-point operations scales as ${N_S}^3 N_\tau$ with possible logarithmic corrections $\sim \log{N_{\tau}}$ from the sparse LU decomposition. Such a mild scaling with $N_\tau$ allows us to enlarge the Euclidean time extent of the lattice and work in the regime where systematic errors produced by the Trotter decomposition are negligible. In terms of scaling with $N_{\tau}$ at fixed $N_S$, the Schur complement solver definitely outperforms the Conjugate Gradient solver, see e.g. Fig.~2 in \cite{Ulybyshev:2018dal}.

Generally, as it was shown in Ref. \cite{Ulybyshev:2018dal}, the Schur complement solver is faster than preconditioned conjugate gradient for moderate lattice sizes up to $N_S=10^3...10^4$, depending on the model. This is one source of speedup. But an even more important source of speedup is that we can typically increase the
integrator stepsize in MD trajectories by at least a factor of $50$ without losing the acceptance rate, if exact fermionic forces  are used. The reason is a much smoother profile of fermionic forces in this case. A comparison of algorithm with pseudofermions (\ref{eq:stoch_det}) and exact (\ref{eq:det_der}) force calculations is shown in Fig. \ref{fig:joint_force}. For these tests, it was possible to achieve an acceptance rate of $\sim0.7$ with exact fermionic forces with an integrator stepsize of $0.2$. Conventional HMC using stochastic representation of determinant  (\ref{eq:stoch_det}) could achieve the same acceptance rate only with the stepsize  $0.0032$. In this case we could decrease the number of steps in MD trajectories by a factor of $60$.

The actual speedup in terms of computer time is approximately half as much, since the iterative computation of the fermionic propagators (\ref{eq:g_off_diag_iter}) makes each integrator step twice as expensive.

\subsection{Observables}
\label{subsec:observables}
 SDW and CDW phases are characterized respectively by the separation of spin and charge between the two triangular sublattices. To study the competition between them in an unbiased way, we introduce order parameters which develop a non-zero expectation value in a finite volume even without any external sources. We use the square of charge and square of spin per sublattice, which are given by   
 \begin{eqnarray}
\label{eq:squarecharge}
 \langle q^2 \rangle  
 =  
 \left\langle \frac{1}{L^4} \left( \sum\limits_{x \in A} \hat{\rho}_x \right)^2 \right\rangle
 +
 \left\langle \frac{1}{L^4} \left( \sum\limits_{x \in B} \hat{\rho}_x \right)^2 \right\rangle , 
\end{eqnarray}
 and
 \begin{eqnarray}
\label{eq:squarespin}
 \langle S^2 \rangle  
 =  
 \left\langle \frac{1}{3L^4} \sum_i\left(
 \sum\limits_{x\in A} \hat{S}_{x,i} \right)^2 \right\rangle 
 + \nonumber \\ +
 \left\langle \frac{1}{3L^4}\sum_i \left( \sum\limits_{x\in B} \hat{S}_{x,i} \right)^2 \right\rangle , 
\end{eqnarray}
where $L$ is the linear lattice size and
\begin{eqnarray}
 \hat{S}_{x,i} = \frac{1}{2} ( \hat
 c^\dag_{x, \uparrow} , \, \hat c^\dag_{x, \downarrow} )  \sigma_i
 \left(
 \begin{array}{c}
  \hat{c}_{x, \uparrow} \\ 
  \hat{c}_{x, \downarrow} \\
 \end{array} 
 \right)~.
\end{eqnarray}
As the sublattices $A$ and $B$ are equivalent, contributions from both are added to improve the signal-to-noise ratio. 

A non-zero value of $\langle S^2 \rangle$ in infinite volume does not unambiguously signal SDW order, as the same observable becomes finite in a ferromagnetic phase. To rule out ferromagnetic order, we also compute the mean squared magnetization  
 \begin{eqnarray}
\label{eq:squaremag}
 \langle m^2 \rangle =
  \left\langle \frac{1}{3 L^4} \sum_i\left( \sum\limits_{x} \hat{S}_{x,i} \right)^2 \right\rangle .
\end{eqnarray}
See Appendix B of Ref. \cite{Buividovich:2018hubb} for expressions for $\langle S^2 \rangle$, $\langle q^2 \rangle$ and $\langle m^2 \rangle$ in terms of fermionic Green functions.

\subsection{Simulation parameters and data analysis}

Using HMC, we simulate graphene sheets with an equal number $L$ of unit cells along each of the crystallographic axes. We simulate lattices with $L=6,12,18,24$. We impose periodic boundary conditions across the borders of rectangular sectors. This choice of geometry corresponds to that used in Ref. \cite{Smekal:1403.3620} but differs from the Born-von K\'arm\'an boundary conditions used in  \cite{Korner:2017qhf}. All results were obtained at temperatures $T=0.125\textrm{eV}=0.046 \kappa$ with $N_\tau=128$, which leads to a time discretization $\delta_{\tau}=0.168 \kappa^{-1}$. This choice is justified by the study of discretization effects in the exponential fermion matrix (\ref{eq:fermionmatrix}) made in \cite{Buividovich:16:4}, where it was shown that the value of squared spin per sublattice (\ref{eq:squarespin}) already stabilizes at this  $\delta_{\tau}$. Thus, we can skip a rather expensive study of the $\delta_{\tau} \rightarrow 0$ limit. We stress here that a similar conclusion does not necessarily follow for other observables, so a convergence with respect to $\delta_{\tau}$ should  carefully be checked in all future work. We choose $\eta=0.9$ as the mixing factor between real and imaginary parts of the Hubbard field introduced in Eq. (\ref{eq:complexint}), which is sufficient to ensure ergodic trajectories \cite{Buividovich:2018hubb,Ulybyshev:1712.02188}. For each produced lattice configuration we compute the full fermionic equal-time Green function $g(x,y)=\langle\hat{a}_x \hat{a}^{\dagger}_y\rangle=M^{-1}_{x,t,y,t}$, which is
  then used to compute all observables. To account for possible autocorrelation effects in our data, we use binning to calculate statistical errors. Typical sample sizes are on the order of several hundreds of independent measurements for a fixed set of parameters. 

\section{Results}
 \label{sec:Results}
 
  We begin with an unbiased study of the competition between CDW and SDW order along the lines of what was recently done for the extended Hubbard model with onsite and nearest-neighbor interactions \cite{Buividovich:2018hubb}. We compute $\sqrt{\langle S^2 \rangle}$ and  $\sqrt{\langle q^2 \rangle}$ for values of $\lambda$ in the range $[1.45 \ldots 1.8]$. We use the square roots of (\ref{eq:squarecharge}) and (\ref{eq:squarespin}) here, as they are characterized by an approach to the infinite volume limit which is linear in $1/L$ to good approximation, which is convenient for extrapolations to the thermodynamic limit. 
 We use linear fits of the form $f(1/L)=a+b\cdot(1/L)$ to lattice sizes $L=12,18,24$ to carry out an $L\to \infty$
 extrapolation. The approach to infinite volume is demonstrated in Fig. \ref{fig:lin_inf_extrapolations}, while the final results is shown in Fig. \ref{fig:inf_extrapolated_miransky}. 
 
 From these results we can immediately conclude that SDW order is favored over CDW order: while the extrapolation of $\sqrt{\langle q^2 \rangle}$ is consistent with zero for any of the coupling strengths considered, $\sqrt{\langle S^2 \rangle}$ develops a non-zero expectation value around $\lambda_c \approx 1.65$. A more precise estimate of $\lambda_c$ will be given below. To rule out a ferromagnetic phase we also measure $\sqrt{\langle m^2 \rangle}$. We find that it is smaller than $\sqrt{\langle S^2 \rangle}$ by an order of magnitude for each parameter set and extrapolates to zero within errors for all cases as well. See Fig. \ref{fig:lin_inf_extrapolations_ferro} for an illustration. 
 
\begin{figure*}[h!tpb]
  \includegraphics[width=0.48\textwidth]{{{extrapolation_plots_full_spin_1}}}
  \includegraphics[width=0.48\textwidth]{{{extrapolation_plots_full_charge_1}}}\\
    \caption{Linear $L\to \infty$ extrapolations of $\sqrt{\langle S^2 \rangle}$ (left) and $\sqrt{\langle q^2 \rangle}$ (right). Fits of the form $f(1/L)=a+b\cdot(1/L)$ to lattice sizes $L=12,18,24$ are shown. \label{fig:lin_inf_extrapolations}}
\end{figure*}

\begin{figure}[h!tpb]
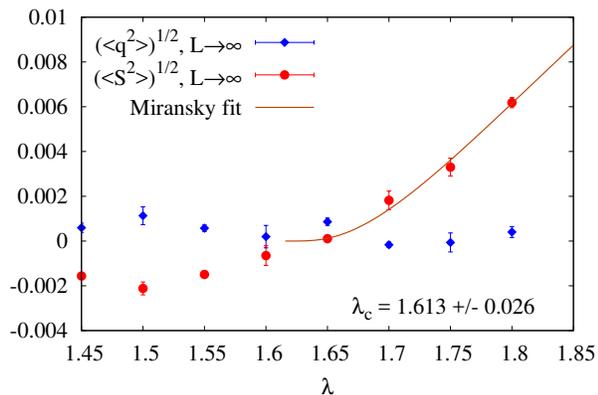

  \includegraphics[width=0.48\textwidth]{{{longrange_inf_miransky}}}\\
    \caption{$L\to \infty$ limit of $\sqrt{\langle S^2 \rangle}$ (red) and $\sqrt{\langle q^2 \rangle}$ (blue). For spin a fit of the Miransky scaling function (\ref{eq:miransky_scaling}) to the data at $\lambda>1.61$ is shown.}
    \label{fig:inf_extrapolated_miransky}
\end{figure}

\begin{figure}[h!tpb]
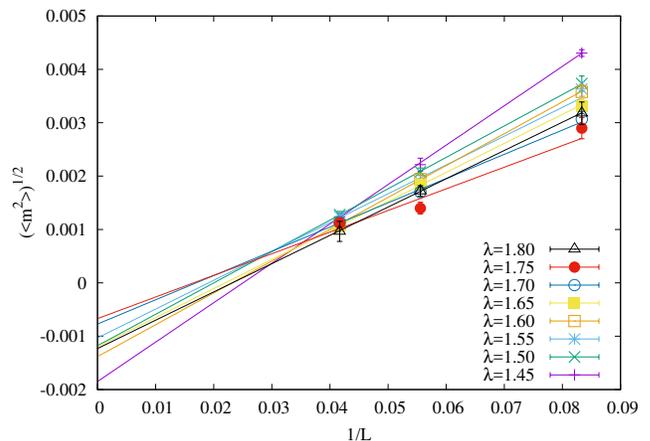

  \includegraphics[width=0.48\textwidth]{{{extrapolation_plots_full_ferro_1}}}\\
      \caption{Linear $L\to \infty$ extrapolations of $\sqrt{\langle m^2 \rangle}$. Fits of the form $f(1/L)=a+b\cdot(1/L)$ to lattice sizes $L=12,18,24$ are shown. For each $L$ and $\lambda$ the signal is weaker by an order of magnitude than both $\sqrt{\langle S^2 \rangle}$ and $\sqrt{\langle q^2 \rangle}$.} \label{fig:lin_inf_extrapolations_ferro}
\end{figure}

To investigate the universal properties of the semimetal-SDW transition, we study the critical scaling of $\langle S^2 \rangle$, which under the assumption of a second-order phase transition should respect
\begin{equation}
\langle S^2 \rangle = L^{-2\beta/\nu}  f(x)\, , \label{eq:critscaling}
\end{equation}
where $f(x)$ is a universal finite-size scaling function and $x = L^{1/\nu} \lr{\lambda - \lambda_c}/\lambda_c$ is the finite-size scaling parameter. Assuming naively that the transition is indeed of second order, we will use Eq. (\ref{eq:critscaling}) to estimate values for $\lambda_c$, the ratio $\beta/\nu$, as well as for the correlation length exponent $\nu$ itself. 

\emph{A priori} the most obvious candidate for the universality class is the $D=3$, $N_f=2$ chiral Heisenberg Gross-Neveu model, based on the hypothesis that the main driving force of the phase transition are onsite interactions (as suggested by RG studies \cite{Herbut2006,Herbut2009,Herbut2009:2,Semenoff2012}). While critical exponents for this class have been obtained in various ways (see e.g. Table I in Ref. \cite{Buividovich:2018hubb} for a summary), it is useful to also examine the Hubbard model (with onsite potential $U$ only) directly here, and obtain a data set which can be used as a point of reference in a one-to-one comparison. This will be the first step of our analysis. In principle the required simulations could be carried out using our HMC code, but for practical reasons\footnote{For the pure Hubbard model BSS is still faster by an order of magnitude than HMC, and remains the method of choice for simulations of contact interactions. HMC is advantageous for long-range interactions, as the number of auxiliary fields is independent of the choice of potential, whereas each interaction term requires an additional field in BSS-QMC.; For a publicly available BSS-QMC code see https://git.physik.uni-wuerzburg.de/ALF.} we choose to produce this data using a GPU implementation of BSS determinantal Quantum-Monte-Carlo \cite{Blankenbecler:81:1} instead, which we will not describe here as the method is widely known (interested readers are referred e.g. to Refs. \cite{Gerlach2017,SciPostPhys.3.2.013}).

\begin{figure*}[h!tpb]
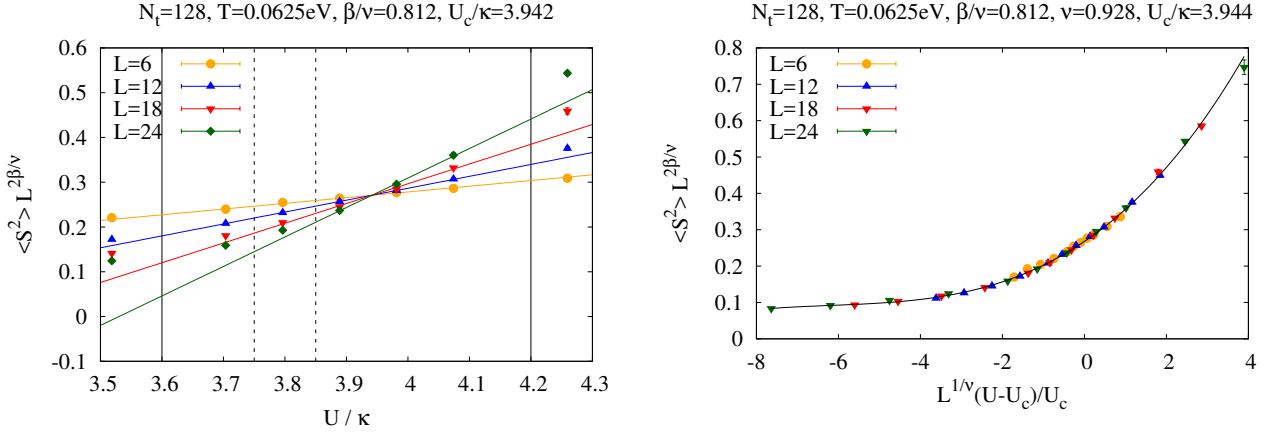

  \includegraphics[width=0.48\textwidth]{{{Sx_scaling_T0.0625_bsshubbard_optimal_withfit_zeromodes}}}
  \includegraphics[width=0.48\textwidth]{{{Sx_collapse_T0.0625_bsshubbard_optimal_withfit_zeromodes_diagnostic}}}\\
    \caption{On the left: $\langle S^2 \rangle L^{2\beta/\nu}$ for the Hubbard model with onsite potential ($U$) only, obtained from BSS-QMC calculations on $L=6,12,18,24$. Solid lines mark boundaries of linear fits to $L=6,12$ data, whereas two dashed lines mark lower bounds of fits to $L=18$ and $L=24$ respectively (upper bounds marked by solid line). Optimized $\beta/\nu$ and $U_c$ are obtained by minimizing enclosed area of all subsets of three lines. On the right: Optimized collapse of $\langle S^2 \rangle L^{2\beta/\nu}$ of the same data onto a universal finite-size scaling function $f(x)$, where $x = L^{1/\nu} (U-U_c)/U_c$. $\beta/\nu$ is fixed to the value obtained in the left panel (both $\nu$ and $U_c$ are optimized as a consistency check).}
    \label{fig:bsshubbard_crit_scaling}
\end{figure*}

Using BSS to simulate the Hubbard model, we obtain data for $\langle S^2 \rangle$ for several values $U$ around the phase transition, which has been estimated to occur at $U_c /\kappa \approx 3.8$ \cite{Assaad:1304.6340}, with all other external parameters matching those of our graphene simulations. To extract both $U_c$ and  $\beta/\nu$ from the data we determine a choice of $\beta/\nu$ for which the $U$-dependent curves of $\langle S^2 \rangle L^{2\beta/\nu}$ for different $L$ intersect in one point. This is done by fitting the data from $L=6,12,18,24$ with linear functions close to the presumed transition point and adjusting $\beta/\nu$ in steps of $0.001$ until the enclosed triangles of all subsets of three lattices are minimized. Note that this requires choosing smaller fit windows on larger lattices, as the curvature of the order paramter grows with system size. Through this method, we find $\beta/\nu \approx 0.8$ and $U_c/\kappa \approx 3.9$ consistent with the previous measurements. 
After fixing the value of $\beta/\nu$ we then obtain $\nu$ by optimizing the collapse of $\langle S^2 \rangle L^{2\beta/\nu}$ onto a universal scaling function, 
by fitting all data points from $L=6,12,18,24$ with a single polynomial function of $x=L^{1/\nu}(U-U_c)/U_c$ (we find that we must use a polynomial of third order) and adjusting $\nu$, also in steps of $0.001$, until the $\chi^2$ per degree of freedom becomes minimal. With $\beta/\nu=0.812$ and $U_c/\kappa=3.942$ this yields $\nu=0.928$.\footnote{Note that the three digits quoted for all results here reflect the resolution used in our optimization procedure and by no means imply a corresponding accuracy.}  As a cross-check $U_c$ is also allowed to shift, yielding an optimal value of $U_c/\kappa=3.944$ for the data collapse. Fig. \ref{fig:bsshubbard_crit_scaling} summarizes these results. 

A few comments are in order here: In principle our predictions for $U_c$ and $\beta/\nu$ should be affected by a systematic uncertainty due to a sensitivity to the windows in which linear fits are applied. We find however that these values are remarkably stable under variations of the fit windows and, quite conservatively, estimate these errors to be $\sim 1\%$ for $U_c$ and $\sim 2\%$ for $\beta/\nu$, which is also most likely larger than our statistical errors. Our results are quite close to the values $\beta/\nu \approx 0.9$ and $U_c /\kappa \approx 3.8$ quoted in Ref. \cite{Assaad:1304.6340}, with the difference likely being due to finite size and temperature effects, which we suspect are the leading source of errors in our case. Likewise, finite size is likely the leading source of uncertainty for $\nu$. If we exclude the $L=6$ lattice and both the $L=6,12$ lattices from the optimized collapse we obtain $\nu=1.037$ and $\nu=1.024$ respectively, suggesting a combined finite-size and statistical error of at least $\sim 6\%$ (note that $\nu$ is slightly larger when only $L=6$ is excluded, suggesting that the we are already close to the thermodynamic limit). We point out here that our values are very much in line with those typically seen in in Monte-Carlo simulations of Hubbard-type models believed to fall into the Gross-Neveu universality class \cite{PhysRevX.6.011029,PhysRevB.91.165108,PhysRevB.90.085146,Assaad:1304.6340}. Renormalization group studies tend to observe slightly larger values, whereby results as large as $\beta/\nu\approx 1.0$ and $\nu\approx 1.2$ have been predicted \cite{Gracey:1801.01320,Knorr:2018,HerbutScherer2017}. 

\begin{figure}[h!tpb]
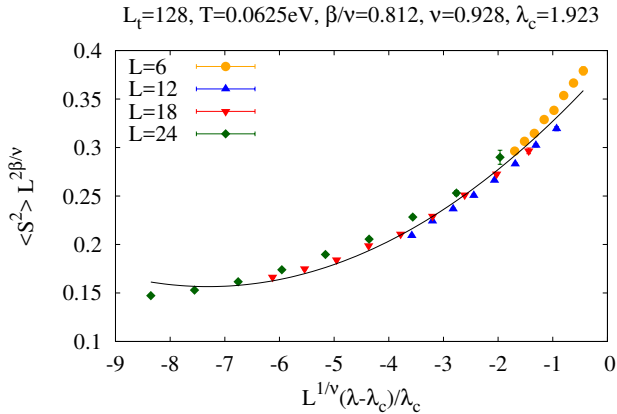

  \includegraphics[width=0.48\textwidth]{{{Sx_collapse_T0.0625_longrange_Hubbard_exponents_withfit_zeromodes_diagnostic}}}\\
      \caption{Attempted collapse of $\langle S^2 \rangle L^{2\beta/\nu}$ from graphene simulations onto a universal finite-size scaling function $f(x)$, where $x=L^{1/\nu}(\lambda-\lambda_c)/\lambda_c$. We fix $\beta/\nu=0.812$ and $\nu=0.928$, corresponding to the optimal values for the pure Hubbard model, and determine $\lambda_c$ by $\chi^2$ optimization. 
      Clearly these choices of critical exponents fail for the realistic potential of graphene.
     } \label{fig:poor_collapse}
\end{figure}

We now turn to the data generated with the realistic potential of graphene. The next logical step is to determine whether we can characterize the critical properties using the same exponents as for the Hubbard model. Thus, we
fix $\beta/\nu=0.812$ and $\nu=0.928$ and test both the intersection of $\langle S^2 \rangle L^{2\beta/\nu}$ for different $L$ and the collapse of data onto a universal function $f(x)$. We find that we can clearly rule out this possibility: Not only do $\langle S^2 \rangle L^{2\beta/\nu}$ intersect nowhere in the region $\lambda\leq 2.0$, contradicting the results shown in Fig. \ref{fig:inf_extrapolated_miransky}, the quality of collapse is also very poor and leads to an unreasonably large estimate of $\lambda_c$. The best possible collapse is shown in Fig. \ref{fig:poor_collapse} while we refrain from even showing any figures for the intersection. 

\begin{figure*}[h!tpb]
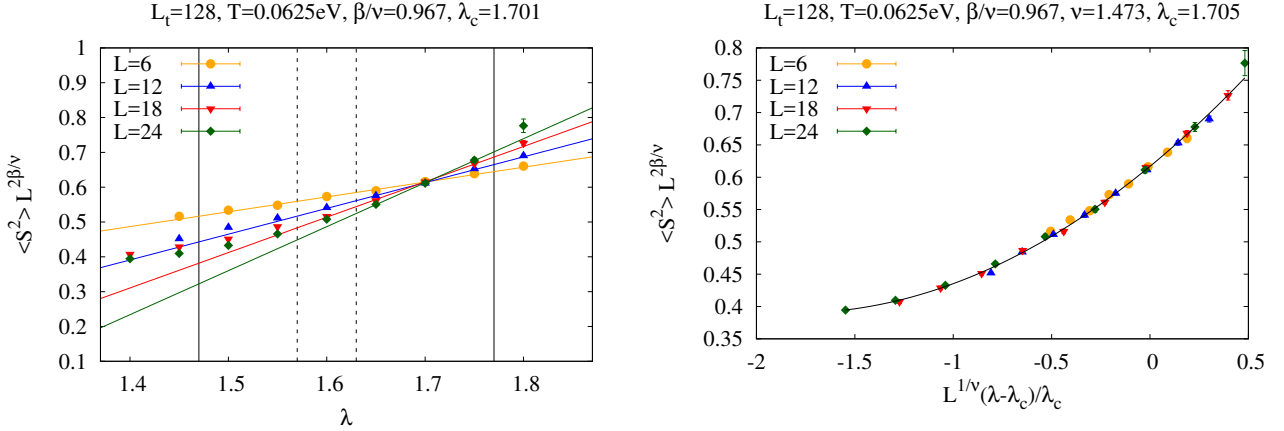

  \includegraphics[width=0.48\textwidth]{{{Sx_scaling_T0.0625_longrange_optimal_withfit_zeromodes}}}
  \includegraphics[width=0.48\textwidth]{{{Sx_collapse_T0.0625_longrange_optimal_withfit_zeromodes_diagnostic}}}\\
    \caption{On the left: $\langle S^2 \rangle L^{2\beta/\nu}$ for graphene with realistic potential as a function of coupling strength $\lambda$ for $L=6,12,18,24$.
    Linear fits were applied in region bounded by vertical lines (dashed lines mark lower bounds of fits to $L=18$ and $L=24$ respectively). Optimized $\beta/\nu$ and $\lambda_c$ were obtained by minimizing the enclosed area of all subsets of three lines. On the right: Optimized collapse of $\langle S^2 \rangle L^{2\beta/\nu}$ onto a universal finite-size scaling function $f(x)$, where $x=L^{1/\nu}(\lambda-\lambda_c)/\lambda_c$. $\beta/\nu$ was fixed to the value obtained in the left panel. $\nu$ and $\lambda_c$ were optimized by
    minimizing the residual 
    $\chi^2$ per degree of freedom of fits of all data points to polynomial functions $f(x)$. An independent consistency check for $\lambda_c$ is obtained. }
    \label{fig:extrapolation_Sx_1}
\end{figure*}

In order to obtain first-principles estimates of $\lambda_c$, $\beta/\nu$ and $\nu$ for graphene we now repeat the same steps as for the Hubbard model: To extract $\lambda_c$ and $\beta/\nu$ from the data we determine a choice of $\beta/\nu$ for which linear fits to $\langle S^2 \rangle L^{2\beta/\nu}$ for different $L$ intersect in one point. Using the resulting $\beta/\nu$ we then determine $\nu$ by optimizing the collapse. In doing so we observe a somewhat odd behavior: It appears that the data, while exhibiting just as small statistical errors for each data point, don't constrain $\beta/\nu$ and $\lambda_c$ nearly as strongly as for the Hubbard model. By choosing different fit windows we can obtain values of $\beta/\nu$ in the range $[0.95,1.0]$ (with the different lines crossing in one point in each case), while $\lambda_c$ shifts in the range $[1.62,1.7]$. The left panel of Fig. \ref{fig:extrapolation_Sx_1} represents our best possible choice of fit windows (with smallest $\chi^2$ for the fits) and leads to $\lambda_c\approx 1.70$ and $\beta/\nu=0.967$. Compared to Fig. \ref{fig:inf_extrapolated_miransky} this estimate for $\lambda_c$ seems slightly too large. 

\begin{figure*}[h!tpb]
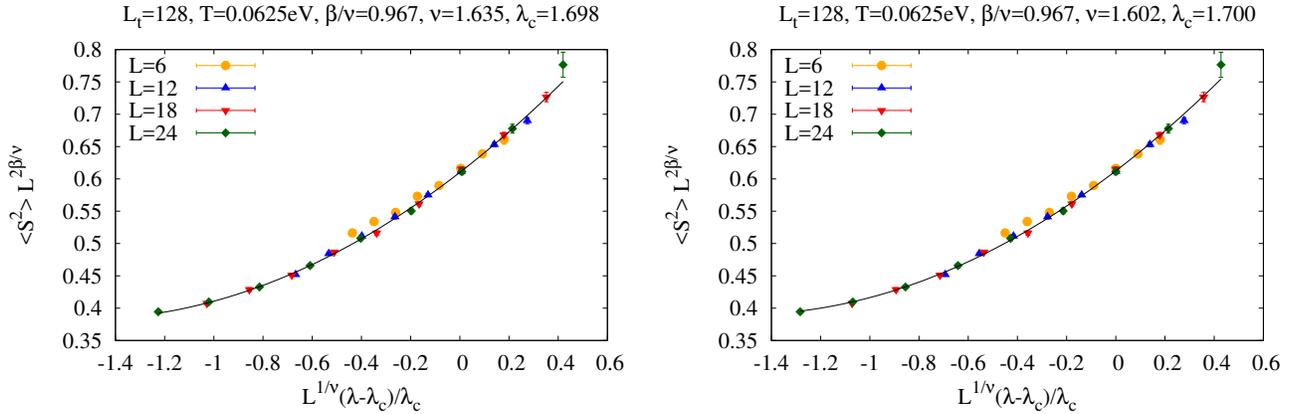

  \includegraphics[width=0.48\textwidth]{{{Sx_collapse_T0.0625_longrange_optimal_withfit_zeromodes_diagnostic_no6}}}
  \includegraphics[width=0.48\textwidth]{{{Sx_collapse_T0.0625_longrange_optimal_withfit_zeromodes_diagnostic_no6_no12}}}\\
    \caption{Optimized collapse of $\langle S^2 \rangle L^{2\beta/\nu}$ for graphene with realistic potential onto a universal finite-size scaling function $f(x)$, $x = L^{1/\nu} \lr{\lambda - \lambda_c}/\lambda_c$ with fixed $\beta/\nu$, when either the $L=6$ data (left panel) or both the $L=6,12$ data (right panel) are excluded from the optimization procedure. Compared to Fig. \ref{fig:extrapolation_Sx_1}, the optimal value of
    $\nu$ slightly increases as the smaller lattices are discarded, while $\lambda_c$ remains stable.}
    \label{fig:exponent_drift}
\end{figure*}

When determining $\nu$ a similarly peculiar observation is made: While an optimized collapse of the $L=6,12,18,24$ data onto a universal function $f(x)$, $x = L^{1/\nu} \, (\lambda - \lambda_c)/\lambda_c$ yields $\nu=1.473$ (right panel of Fig. \ref{fig:extrapolation_Sx_1}), we find that the optimal value of $\nu$ increases slightly when the $L=6$ lattice is excluded from the fit, resulting in the optimal choice of $\nu=1.635$ for fits to $L=12,18,24$ and $\nu=1.602$ for $L=18,24$ (see Fig. \ref{fig:exponent_drift}). While the spread of these numbers suggests that the combined statistical and finite-size error can be as large as $\sim 10 \%$, the observation that the optimal value moves further away from that of the Hubbard model on larger lattices is unexpected.
We point out here that it was sufficient to model the universal scaling function with a quadratic polynomial for the case of graphene.

One could still be inclined to interpret these results in terms of Gross-Neveu scaling: As our value of $\beta/\nu$ for graphene is closer to the RG result ($\beta/\nu\approx 1.0$) than for the Hubbard model, one might speculate that non-universal corrections to scaling are more severe in the case of a pure onsite potential for some reason, that RG exponents are closer to the true values than MC predictions, and that our results with the long-range potential reflect the true universal behavior more closely. Without commenting on the plausiblity of this scenario we note that our smallest prediction for $\nu$ is still $\sim 20\%$ larger than the RG result ($\nu\approx 1.2$), however. 

\begin{figure}[h!tpb]
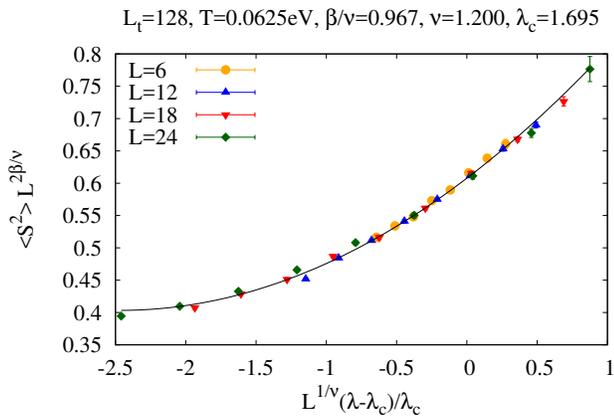

   \includegraphics[width=0.48\textwidth]{{{Sx_collapse_T0.0625_longrange_fixednu_withfit_zeromodes_diagnostic}}}\\
   \caption{Collapse of $\langle S^2 \rangle L^{2\beta/\nu}$ onto a universal finite-size scaling function with fixed $\beta/\nu=1.0$ and $\nu=1.2$ ($\sim$ RG prediction for $D=3$, $N_f=2$ chiral Heisenberg Gross-Neveu class \cite{Gracey:1801.01320,Knorr:2018,HerbutScherer2017}). Only $\lambda_c$ is optimized. Quality of fit is significantly lower than with variable exponents (see Fig. \ref{fig:collapse_chisquare}).} \label{fig:fixednu_collapse}
\end{figure}

\begin{figure*}[h!tpb]
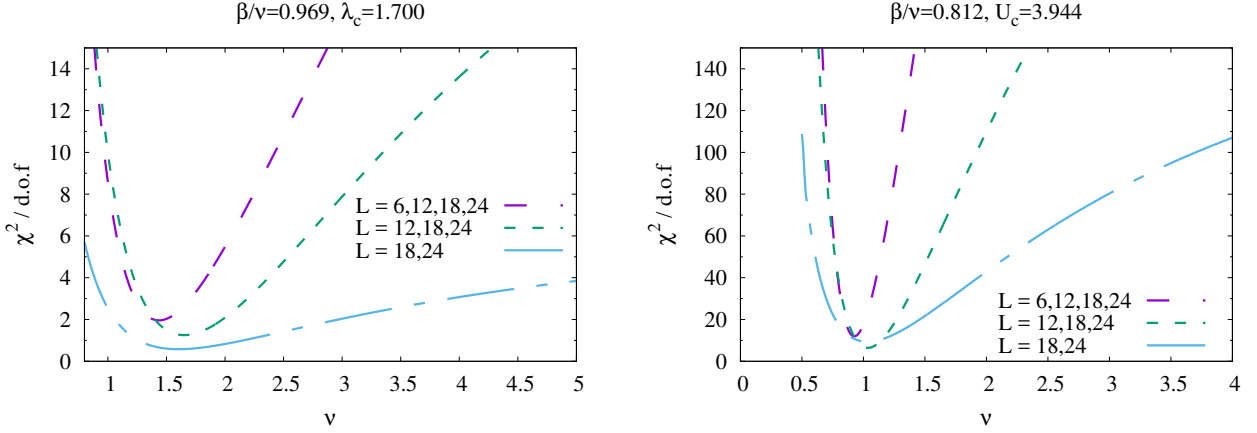

  \includegraphics[width=0.48\textwidth]{{{collapse_chisquare_graphene}}}
  \includegraphics[width=0.48\textwidth]{{{collapse_chisquare_bsshubbard}}}
  \caption{$\chi^2$ per degree of freedom for collapse of $\langle S^2 \rangle L^{2\beta/\nu}$ onto a universal finite-size scaling function for graphene (left) and Hubbard model (right) with different choices of $\nu$ for different sets of lattice sizes. $\beta/\nu$ and critical coupling strength are fixed to their optimal values. }
  \label{fig:collapse_chisquare}
\end{figure*}

To get a sense of how significant this deviation is (in principle a visually only slightly less convincing collapse can be obtained with $\nu=1.2$, see Fig. \ref{fig:fixednu_collapse}) we do the following: After fixing $\beta/\nu=0.967$ and $\lambda_c=1.7$ we shift $\nu$ in the range $[0.5,5.0]$ and obtain the $\chi^2$ per degree of freedom resulting from fits of quadratic polynomials to $\langle S^2 \rangle L^{2\beta/\nu}$ as a function of $x = L^{1/\nu}(\lambda-\lambda_c)/\lambda$ for each choice of $\nu$.
We do this for the full set of lattice sizes $L=6,12,18,24$ and for the cases where $L=6$ and $L=6,12$ are ignored. A similar procedure is repeated for the Hubbard model (with $\beta/\nu=0.812$, $U_c=3.944$ and third order polynomials) for comparison. The results are shown in Fig. \ref{fig:collapse_chisquare}. 

The first thing we can say is that, on a quantitative level, $\nu=1.2$ is clearly disfavored for graphene. While for the $L=6,12,18,24$ lattices the $\chi^2/\textrm{d.o.f}$ is about $1.7$ times larger at $\nu=1.2$ compared the optimum, the ratio grows even larger as small lattices are excluded, up to about $2.0$ on lattice sizes $L=18,24$. 

An interesting observation is that the $\chi^2$ curves for both the Hubbard model and graphene become flatter at large $\nu$ as small lattice sizes are excluded. To some degree this is not surprising, as a smaller number of data points places weaker constraints on the parameters of the polynomials used to model the universal scaling function. What is striking however is the substantial difference between graphene and the Hubbard model. For the Hubbard model, the scale of the y-axis is larger by an order of magnitude, despite the fact that higher order polynomials were used, reflecting the fact that critical exponents are constrained much stronger by the data.  For graphene on the other hand, it appears as if the curves very quickly converge to a completely flat curve with $\chi^2 \sim 1$ for $\nu\gtrsim2.0$ on larger volumes, suggesting that for large systems any sufficiently large choice of $\nu$ will work equally well. This is not at all what one expects for a regular second order phase transition.

To conclude our analysis, we conservatively state that the numerical data for the long-range interaction potential is quite different from the data obtained for purely on-site interaction, and is hardly consistent with the Gross-Neveu universality class. We conjecture that instead our results could be interpreted as signatures of Miransky scaling. 

A CPT is known to occur in quantum electrodynamics (QED) with massless fermions, both in $3+1$ and $2+1$ dimensions, with the number of fermion flavors being the control parameter in the later case \cite{PhysRevD.58.085023,PhysRevLett.75.2081,PhysRevLett.60.2575,Braun:2014wja}.\footnote{QED$_{2+1}$ has been considered as a model for a similar transition believed to occur in $SU(N)$ gauge theories (see e.g. Refs. \cite{BANKS1982189,PhysRevD.84.125006,PhysRevLett.100.171607,Braun:2009ns,Gies:2005as} and references therein). In QED$_{2+1}$ one speaks of a ``pseudo-conformal'' transition as conformal symmetry is explicitly broken by a dimensionful coupling constant but the theory nevertheless exhibits an effective low-energy scale invariance.} The proper low-energy effective theory of graphene (``reduced QED$_4$''), combines features of both, as electron motion is restricted to a plane but photons can propagate in the three dimensional bulk (one also takes the smallness of the Fermi velocity into account, which leads to effectively instantaneous Coulomb interactions with an exact conformal symmetry). In this theory, a chiral phase transition exhibiting Miransky scaling\footnote{A sufficiently strong four-fermion coupling (corresponding roughly to onsite interactions in the microscopic theory) can change the transition to one of second order \cite{Gamayun:2010}.} has been demonstrated by solving Dyson-Schwinger equations \cite{PhysRevB.66.045108, KHVESHCHENKO2004323}. In Ref. \cite{Gamayun:2010} it was argued that this CPT formally corresponds to the limit $\beta, \nu\to\infty$, $\delta=1$ of a second order transition and that the usual hyperscaling relations may apply. In our case $d=2$ (where $d$ is the number of spatial dimensions) and with $\delta=1$ the relation 
\begin{equation}
\frac{\beta}{\nu}=\frac{d}{\delta+1}
\end{equation}
would thus lead to $\beta/\nu=1$, which agrees with our estimate for the optimal value at a level of about  $3 \%$  and is thus well within the present errors.  

For QED$_{2+1}$\cite{PhysRevD.68.025017,PhysRevB.79.064513} and reduced QED$_4$ \cite{PhysRevB.79.205429} it is well known that that the CPT exhibits a strong sensitivity to an infrared cutoff. For many-flavor QCD it was shown that Miransky scaling receives powerlaw corrections from an infrared RG fixed point of the gauge coupling \cite{PhysRevD.84.034045}.  
It is thus reasonable to assume that finite-size effects mimic a second order phase transition, and that  $\beta,\nu\to\infty$ as the thermodynamic limit is approached, with their ratio being fixed by the hyperscaling relation. The slight drift of the exponent $\nu$ towards larger values observed above could be interpreted in these terms as can be the relative insensitivity of the quality of collapse towards further increases of $\nu$ on large lattices. 

 \begin{figure*}[h!tpb]
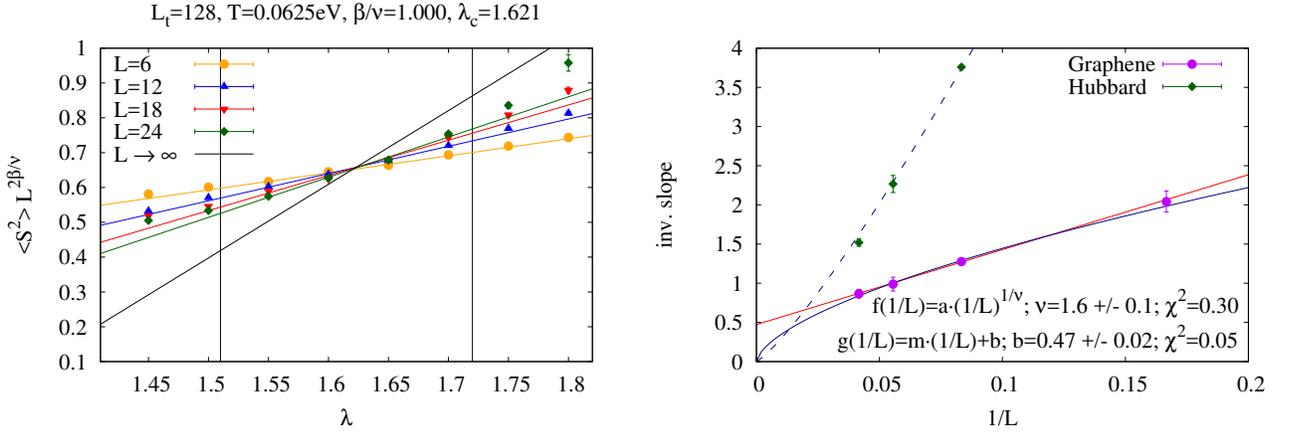

  \includegraphics[width=0.48\textwidth]{{{Sx_scaling_T0.0625_longrange_optimal_withfit_zeromodes_miransky}}}
  \includegraphics[width=0.48\textwidth]{{{scaling_fit_parameters_combined}}}\\
    \caption{On the left: Finite size scaling of $\langle S^2 \rangle$ for graphene with fixed $\beta/\nu=1.0$.  Linear fits are applied in the window marked by vertical lines. On the right: Extrapolation of the inverse slope of the fits to the  thermodynamic limit using models $g(1/L)=m\cdot(1/L)+b$ and $f(1/L)=a\cdot(1/L)^{1/\nu}$ respectively (where $\nu$ is treated as a free parameter). The linear model $g(\cdot)$ (with smaller $\chi^2$) predicts that the slope remains finite for $L\to \infty$, suggesting that the data may collapse onto a universal function without rescaling the coupling constant. Data points for the Hubbard model together with a power-law model curve are shown for comparison.   }\label{fig:miransky_intersection}
\end{figure*}

As a final test we therefore study the finite-size scaling properties of 
$\langle S^2 \rangle$ for graphene with fixed $\beta/\nu=1.0$. We find that linear fits to $\langle S^2 \rangle L^{2\beta/\nu}$ with $L=6,12,18,24$ cross in one point to good precision if we choose the fit windows as $\lambda=[1.525;1.725]$ (see left panel of Fig. \ref{fig:miransky_intersection}) and thereby estimate $\lambda_c\approx 1.62$, which is substantially lower than the result $\lambda_c\approx 1.70$ obtained with $\beta/\nu=0.967$ but falls much closer to the location one would expect, judging by the $L\to \infty$ extrapolated results for $\sqrt{\langle S^2 \rangle}$ shown in Fig.  \ref{fig:inf_extrapolated_miransky}: By fitting the Miransky scaling function
 \begin{equation}
\sigma(\lambda)=a\exp\left(\frac{-b} {\sqrt{\lambda - \lambda_c}} \right)~,
\label{eq:miransky_scaling}
\end{equation}
as appropriate for reduced QED$_4$ (see e.g. Ref. \cite{Gamayun:2010}) to $\sqrt{\langle S^2 \rangle}|_{L=\infty}$ (which works extremely well), we can independently estimate $\lambda_c=1.61 \pm 0.02$ which agrees with our prediction from finite-size scaling to very good precision. This agreement can be seen as evidence in favor of the CPT scenario.

Perhaps more importantly, however, we note that the slopes in the intersection plot in the left panel of Fig.~\ref{fig:miransky_intersection} do not appear to increase towards infinity with increasing volumes as they should for a second-order phase transition. In fact, the solid black line in this plot represents our infinite-volume extrapolation which is clearly not vertical. If true, there would certainly be no way that this could ever happen in a second-order phase transition for any finite value of $\nu $, i.e.~without rescaling the reduced control parameter by $L^{1/\nu}$ along the abscissa.  

To investigate this somewhat more carefully, we plot the inverse slopes of our linear fits to $\langle S^2 \rangle L^{2\beta/\nu}$ from the intersection plots over $1/L$ in the right panel of Fig.~\ref{fig:miransky_intersection}: 

While the Hubbard model data once again shows the expected behavior, the inverse slopes for graphene obtained from the left panel in this figure (with $\beta/\nu=1.0$) are well described by a linear model fit to
\[
g(1/L)=m\cdot(1/L)+b \quad\mbox{with}\quad  b=0.47 \pm 0.02\, . 
\]
This non-zero intercept $b$ then provides our best numerical evidence of a finite slope in the infinite-volume limit and hence of $\beta,\nu \to \infty$ as CPT characteristics. 

As mentioned, this also implies that on sufficiently large lattices a collapse onto a universal finite-size scaling function occurs for $x=(\lambda-\lambda_c)/\lambda_c$ alone, i.e.~without rescaling the reduced coupling by the factor $L^{1/\nu}$, which one expects if $\nu\to\infty$, but which also explains the difficulty in determining a stable value for $\beta/\nu $ from intersection points on larger lattices where the curves do not intersect anymore when the data collapses all by itself in an ever growing region around $\lambda_c$.   

For completeness we close this section with adding that a behavior as expected for a second-order phase transition, modelling the inverse slopes with $f(1/L)=a\cdot(1/L)^{1/\nu}$ without intercept, can also be used to fit the graphene data, resulting in $\nu=1.6 \pm 0.1$, but yielding a $\chi^2$/d.o.f. which is larger by about a factor of six as compared to the linear fit described above.\footnote{Note that $\chi^2$ values are $< 1$ for both models here, which reflects the fact that the slopes and their error bars are themselves the results of linear fits.} That such second-order fits work as well, on sufficiently small lattices, might rather be a manifestation of the difficulty in distinguishing CPT behavior from second-order scaling in finite volumes. 

The direct comparison between Hubbard model and graphene data, however, provides quite compelling evidence that the second-order scaling, which works very well for the former, gets increasingly difficult to maintain with increasing volumes in the case of graphene with long-range interactions where the CPT scenario appears to provide the much more natural explanation.

\section{Conclusion and Outlook}
\label{sec:Conclusion}

In this work we carried out a detailed unbiased study of the competition of SDW and CDW orders in graphene, with a realistic two-body potential that includes an unscreened long-range Coulomb tail. Using state of the art Hybrid-Monte-Carlo simulations we were able to determine that the potential must be uniformly rescaled by roughly a factor of $\sim 1.6$ to induce a semimetal-SDW phase transition, corresponding to a critical value of effective fine-structure constant $\alpha_c \approx 3.5$. This is substantially larger than the value $\alpha_c \approx 3.1$ predicted by previous studies \cite{Buividovich:13:5,Smekal:1403.3620,Assaad:1505.04188} which were much more strongly affected by discretization artifacts, finite-size effects and larger temperatures. We find no evidence for CDW order, confirming that SDW is the preferred phase as predicted by renormalization group analysis  \cite{Herbut2006,Herbut2009,Herbut2009:2} and strong-coupling expansion \cite{Semenoff2012}. 

A careful study of the critical properties suggested that the expected $D=3$, $N_f=2$ chiral Heisenberg Gross-Neveu universality class, as expected for the corresponding Hubbard model, is unlikely to apply to the semimetal-insulator transition in graphene with long-range Coulomb interactions. Our lower-bound estimates for the correlation-length exponent $\nu$ are significantly larger than the largest values predicted by RG studies for this class. A direct comparison with a data set produced for the Hubbard model with on-site interactions only also ruled out with high confidence that both systems can be described by a common set of critical exponents, demonstrating clearly that the long-range part of the inter-electron potential plays an important role in non-perturbative many-body physics of graphene. 

In studying the finite-size scaling of the squared spin per sublattice an unexpected property of $\nu$ was observed: The optimal choice, which produces the best possible collapse of data from different lattice sizes onto a universal finite-size scaling function, seems to drift slightly towards larger values as smaller lattices are disregarded during the analysis, instead of converging towards the RG predictions as one would expect if the difference was due to corrections to scaling. Further investigations revealed that constraints on $\nu$ become weaker on large lattices, such that $\nu$ can be increased without affecting the quality of collapse substantially. This stands in stark contrast to the Hubbard model, where critical exponents are constrained much tighter by the data. Furthermore, we obtained evidence that a collapse may occur naturally in infinite volume, without the need for rescaling the reduced coupling by a factor $L^{1/\nu}$, which formally corresponds to the limit $\nu \to \infty$.

We have proposed that the observed behavior can be explained in terms of a conformal phase transition, exhibiting exponential ``Miransky'' scaling, which is predicted for $2+1$-dimensional Dirac fermions with bare Coulomb interactions
by Dyson-Schwinger studies \cite{PhysRevB.66.045108, KHVESHCHENKO2004323}, and power-law corrections that mimic a second order phase transition caused by finite size effects \cite{PhysRevD.68.025017,PhysRevB.79.064513,PhysRevB.79.205429,PhysRevD.84.034045}. A formal hyperscaling relation between the exponents $\beta$ and $\nu$ seems to be fulfilled in good approximation. 

Let us note that the phase transitions to CDW and SDW ordered states and an infinite-order conformal phase transition are basically the only theoretical predictions for graphene with long-range interactions of which we are aware, and our numerical analysis is intended to find the most likely scenario out of these three. This turns out to be the conformal phase transition scenario. It can be that our data is also consistent with some other exotic scenario which has not been studied so far.

On the technical side, we have described a variant of HMC which achieves a substantial performance improvement by using exact fermionic forces and avoiding the use of pseudofermions.

An obvious direction for future studies is to push the simulations towards even larger system sizes. The $L=24$ lattices studied in this work represent the largest systems which are feasible with our current computational resources. Repeating the finite-size scaling analysis with $L=24,30,36$, to test whether the trend of growing correlation length exponent $\nu$ continues, would be extremely beneficial and should become feasible in the near future. The
infinite volume extrapolations shown in Figs.
 \ref{fig:lin_inf_extrapolations},
\ref{fig:inf_extrapolated_miransky}
 and \ref{fig:lin_inf_extrapolations_ferro}
 as of now use only three points and would also be greatly 
 improved by including additional lattice sizes.
Another possibility is to study in detail how the order of the phase transition depends on the balance of short- and long-range parts of the potential. This could guide experimental efforts to induce a conformal phase transition in real graphene samples through techniques such as mechanical strain \cite{Assaad:1505.04188,PhysRevB.96.155114}.

\medskip

\begin{acknowledgments}
P.B.\ is supported by a Heisenberg Fellowship from the the Deutsche Forschungsgemeinschaft  (DFG), grant BU 2626/3-1. M.U.~is also supported by the DFG under grant AS120/14-1. D.S.~and L.v.S.~ are supported by the Helmholtz International Center for FAIR within the LOEWE initiative of the State of Hesse. Calculations were carried out on GPU clusters at the Universities of Giessen and Regensburg and on the JUWELS system at the J\"ulich Supercomputing Centre. We thank  F.~Assaad, Ch.~Fischer and
B.-J.~Schaefer for helpful discussions.
\end{acknowledgments}

\bibliographystyle{apsrev4-1}
\bibliography{Buividovich,Smith}

\end{document}